\documentstyle[prl,aps,floats,epsf]{revtex}
\begin{document}
\draft 
\wideabs{
\title{Chemical Potential Shift in Nd$_{2-x}$Ce$_{x}$CuO$_{4}$: Contrasting Behaviors of the Electron- and Hole-Doped Cuprates}
\author{N.~Harima, J.~Matsuno, and A.~Fujimori}
\address{Department of Physics and Department of Complexity Science
  and Engineering, University of Tokyo, Bunkyo-ku, Tokyo 113-0033, Japan}
\author{Y.~Onose, Y.~Taguchi, and Y.~Tokura}
\address{Department of Applied Physics, University of Tokyo, Bunkyo-ku,
 Tokyo 113-8656, Japan}
\date{\today}
\maketitle

\begin{abstract}
 We have studied the chemical potential shift in the electron-doped 
 superconductor Nd$_{2-x}$Ce$_{x}$CuO$_{4}$ 
 by precise measurements of core-level photoemission spectra. The result shows 
 that the chemical potential monotonously increases with electron
 doping, quite differently from La$_{2-x}$Sr$_{x}$CuO$_{4}$, 
 where the shift is suppressed in the
 underdoped region. 
 If the suppression of the shift in La$_{2-x}$Sr$_{x}$CuO$_{4}$ 
 is attributed to strong stripe fluctuations, 
 the monotonous increase of the chemical potential is consistent with 
 the absence of stripe fluctuations in Nd$_{2-x}$Ce$_{x}$CuO$_{4}$. 
The chemical potential jump
 between Nd$_{2}$CuO$_{4}$ and La$_{2}$CuO$_{4}$ is found to
 be much smaller than the optical band gaps.

\end{abstract}
}

\narrowtext
High-$T_c$ superconductivity occurs when the parent antiferromagnetic 
(AF) insulator with the CuO$_2$ plane is doped with holes or electrons. 
In the {\it p}-type materials, the long-range AF order vanishes for a slight 
amount of hole doping 
 whereas in the {\it n}-type materials,
the AF order persists up to a high doping concentration 
of ${\sim}$0.14 electrons per Cu and the superconducting (SC) doping range 
is much narrower~\cite{Tokura}. 
The {\it p}-type materials show $T$-linear in-plane electrical 
resistivity~\cite{Takagi} and split neutron peaks around ${\bf q} = (\pi,\pi)$
indicating incommensurate spin fluctuations~\cite{LSCO_neu1} 
whereas the {\it n}-type materials show
$T^2$ dependence of the in-plane resistivity 
\cite{NCCO_res} and $(\pi,\pi)$ commensurate 
spin fluctuations \cite{NCCO_neu}. 
In order to elucidate the mechanism of high-$T_c$ superconductivity,
it is very important to clarify the origin of the 
similarities and the differences between the {\it p}-type 
and the {\it n}-type materials.

In this Letter, we report on a 
study of the chemical potential shift in 
Nd$_{2-x}$Ce$_{x}$CuO$_{4}$ (NCCO) as a function of doped 
electron concentration. The shift can be deduced from the core-level 
shifts in photoemission spectra because the binding energy of 
each core level is measured relative to 
the chemical potential $\mu$. 
In a previous study~\cite{LSCO_mu}, 
we found that in La$_{2-x}$Sr$_{x}$CuO$_{4}$ (LSCO) 
the chemical potential shift is unusually suppressed in the underdoped 
region and attributed this observation 
to the strong stripe fluctuations which exist in this system.
As for the chemical potential jump between La$_2$CuO$_4$ and 
Nd$_2$CuO$_4$, which would represent the band gap of the parent insulator, 
it was estimated to be at most 300 meV in previous valence-band photoemission
studies~\cite{NCCO_AIPES1,NCCO_AIPES2},
which is much smaller than the 1.5--2.0 eV charge-transfer (CT)  
gap of the parent insulator estimated from 
optical studies \cite{Uchida}. 

High-quality single crystals of NCCO ($x=$ 0, 0.05, 0.125 and 0.15)
were grown by the traveling-solvent floating-zone method as 
described elsewhere~\cite{Onose}. 
Uncertainties in the Ce concentration were ${\pm}0.01$.
For $x=0.15$, both as-grown and reduced samples were measured
while for the other compositions only as-grown samples were measured.
The as-grown samples were
all antiferromagnetic and did not show superconductivity.
Only the $x=0.15$ sample showed superconductivity after reduction 
in an Ar atmosphere and its $T_c$ was ${\sim}25$ K.

X-ray photoemission spectroscopy (XPS) measurements were performed using 
both the Mg $K{\alpha}$ (${\it h}{\nu} = 1253.6~$eV) and 
Al $K{\alpha}$ (${\it h}{\nu} = 1486.6~$eV) lines and a hemispherical 
analyzer. 
All the spectra were taken at liquid-nitrogen temperature 
(${\sim}80$~K) within $40$ minutes after scraping.
We did not observe a shoulder on the higher binding
energy side of each O 1{\it s} peak, indicating 
the high quality of the sample surfaces free from degradation.
Although the energy resolution was about $0.8$ eV for both $K{\alpha}$ lines,
we could determine the core-level shifts with an accuracy of about
${\pm}50$ meV because most of the spectral line shapes did not
change with $x$.
In XPS measurements, a high voltage of $>1$ kV is used to decelerate
photoelectrons, and it is usually difficult to stabilize
the high voltage with the accuracy of $\ll 100$ meV.
In order to overcome this difficulty,
we directly monitored the voltage applied to the outer hemisphere and the 
retarding fringe,
and confirmed that the uncertainty could be reduced to less than $10$ meV.
To eliminate other unexpected causes of errors,
we measured the $x=0.05$ sample as a reference 
just after the measurement of each sample. 

\begin{figure}[htb]
\begin{center}
\epsfxsize=80mm
\epsfbox{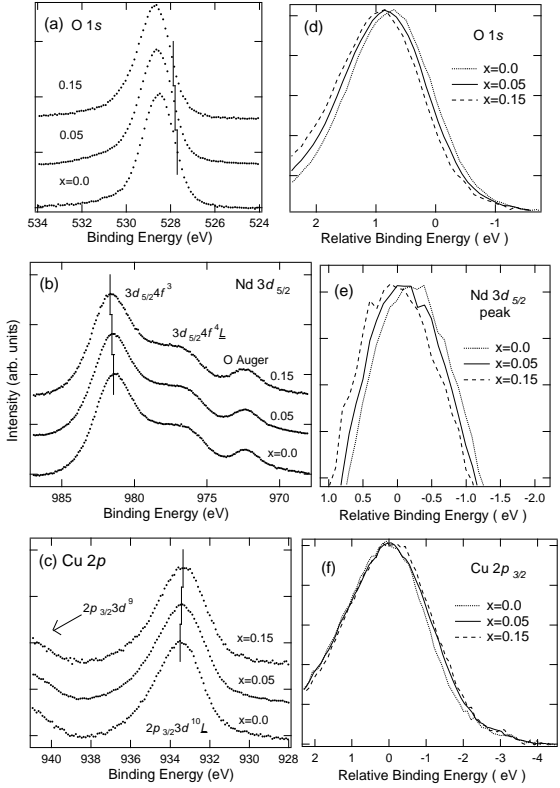}
\caption{Core-level photoemission spectra of 
Nd$_{2-x}$Ce$_x$CuO$_4$ taken with the Al $K{\alpha}$ line. 
(a),(d): O 1{\it s};  (b),(e): Nd 3{\it d};  (c),(f): Cu 2{\it p}.
Integral background has been subtracted and 
the intensity has been normalized to the peak height.}
\end{center}
\end{figure}
Figure 1 shows the XPS spectra of the O 1{\it s}, Nd 3$d_{5/2}$ and 
Cu 2$p_{3/2}$ core levels taken with the Al $K{\alpha}$ line.
Here, the integral background has been subtracted and 
the intensity has been normalized to the peak height \cite{rem}.
The Nd 3$d_{5/2}$ spectra are composed of the 3$d_{5/2}$4$f^4$\b{\it L} and
3$d_{5/2}$4$f^3$ final-state components, where \b{\it L} denotes a ligand hole,
and O {\it KLL} Auger signals overlap them. 
The Cu 2$p_{3/2}$ spectra are composed of the 2$p_{3/2}$3$d^{10}$\b{\it L} and
2$p_{3/2}$3$d^{9}$ components, 
but only the 2$p_{3/2}$3$d^{10}$\b{\it L} peaks are shown in the figure.
One can see the obvious doping dependent shifts of O $1s$ and Nd $3d$  
core levels
from both the displaced and overlayerd plots in Fig. 1.
To deduce the amount of the core-level shifts reliably, 
we used the peak position 
for the Nd 3{\it d} spectra and the mid point of the lower binding energy 
slope for the O 1{\it s} spectra. We used the mid-point position 
rather than the peak position for O $1s$ 
because the line shape on the higher binding energy side of the O 1{\it s} peak
was sensitive to a slight surface degradation
or contamination. The Cu 2{\it p} core-level line shape was 
not identical between different $x$'s, and becomes broader as $x$ increases.
This is because the doped electrons in the CuO$_{2}$ plane produce
Cu$^{1+}$ sites on the otherwise Cu$^{2+}$ background, which yield 
an overlapping chemically shifted component located
on the lower binding side of the Cu$^{2+}$ peak.
Therefore, it was difficult to uniquely determine the shift of the
Cu 2{\it p} core level and we only take its peak positions in the following. 

Figure 2 shows the binding energy shift 
of each core level relative to the as-grown $x=0.05$ sample. 
Here, we have assumed that the change of the electron concentration 
caused by the oxygen reduction
was ${\sim}0.04$ per Cu (oxygen reduction being ${\sim}0.02$) as reported 
previously~\cite{oxygen_reduction}.
One can see that the Nd 3{\it d} and O 1{\it s} levels move toward higher binding 
energies with electron doping.
The shift of Cu 2{\it p} is defined by the shift of 
the peak position,
and is in the opposite direction to Nd 3{\it d} and O 1{\it s}
because of the Cu$^{1+}$ components mentioned above.
We also measured the shifts of the core levels using 
the Mg $K{\alpha}$ line and almost the same results were obtained
as shown in Fig. 2. 
\begin{figure}[htb]
\begin{center}
\epsfxsize=80mm
\epsfbox{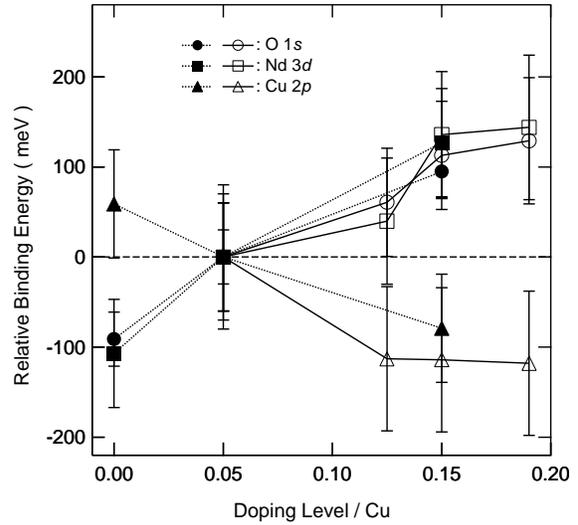}
\caption{Binding energy shift of each core level 
relative to the $x=0.05$ sample. Open and filled symbols are data
taken with the Mg $K{\alpha}$ and Al $K{\alpha}$ radiation, respectively.}
\end{center}
\end{figure}

While the shift of the chemical potential changes
the core-level binding energy, there is another factor
that could affect the binding energy, that is, the change in the Madelung 
potential due to Ce$^{4+}$ 
substitution for Nd$^{3+}$.
However, the identical shifts of the 
O 1{\it s} and Nd 3{\it d} core levels indicate
that the change in the 
Madelung potential has negligible affects on the core-level shifts
because it would shift the core levels of the
O$^{2-}$ anion and Nd$^{3+}$ cation in the opposite directions.
Moreover, as the shifts of the O $1s$ and Nd $3d$ core levels 
toward higher binding energies with electron doping are opposite
to what would be expcted from increasing core-hole screening capability 
with $x$, 
excluding the core-hole screening mechanism as the main cause of 
the core-level shifts.  
Therefore, we conclude that the shifts of the O $1{\it s}$ 
and Nd $3{\it d}$ core levels 
are largely due to the chemical potential shift ${\Delta}{\mu}$.
We have evaluated ${\Delta}{\mu}$ in NCCO by taking the average of
the shifts of the two core levels.

Figure 3(a) shows ${\Delta}{\mu}$ in NCCO 
as well as ${\Delta}{\mu}$ in LSCO~\cite{LSCO_mu}
as a function of electron or hole carrier concentration.
In order to obtain the jump in $\mu$ 
between Nd$_{2}$CuO$_4$ and La$_{2}$CuO$_4$, 
we also measured the O $1{\it s}$ and Cu $2{\it p}$ levels in LSCO as
shown in Fig.~4, and found
that the O $1{\it s}$ and Cu $2{\it p}$ levels in 
Nd$_{2}$CuO$_{4}$ lie at ${\sim}150$ meV and ${\sim}400$ meV higher 
binding energies than those in La$_{2}$CuO$_{4}$, respectively. 
The fact that the observed jump is different between O 1{\it s} and Cu 2{\it p}
is not surprising because Nd$_{2}$CuO$_{4}$ and La$_{2}$CuO$_{4}$ are
different materials with different crystal structures.
Thus the chemical potential jump between Nd$_{2}$CuO$_{4}$ and 
La$_{2}$CuO$_{4}$ cannot be uniquely determined from those data but it 
should be much smaller than the CT gap of about $1.5$ eV for Nd$_{2}$CuO$_{4}$ 
and $2.0$ eV for La$_{2}$CuO$_{4}$ estimated from the optical 
measurements \cite{Uchida}.
This small jump is in accordance with the early valence-band photoemission
studies of LSCO and NCCO~\cite{NCCO_AIPES1,NCCO_AIPES2}.
\begin{figure}[hb]
\begin{center}
\epsfxsize=80mm
\epsfbox{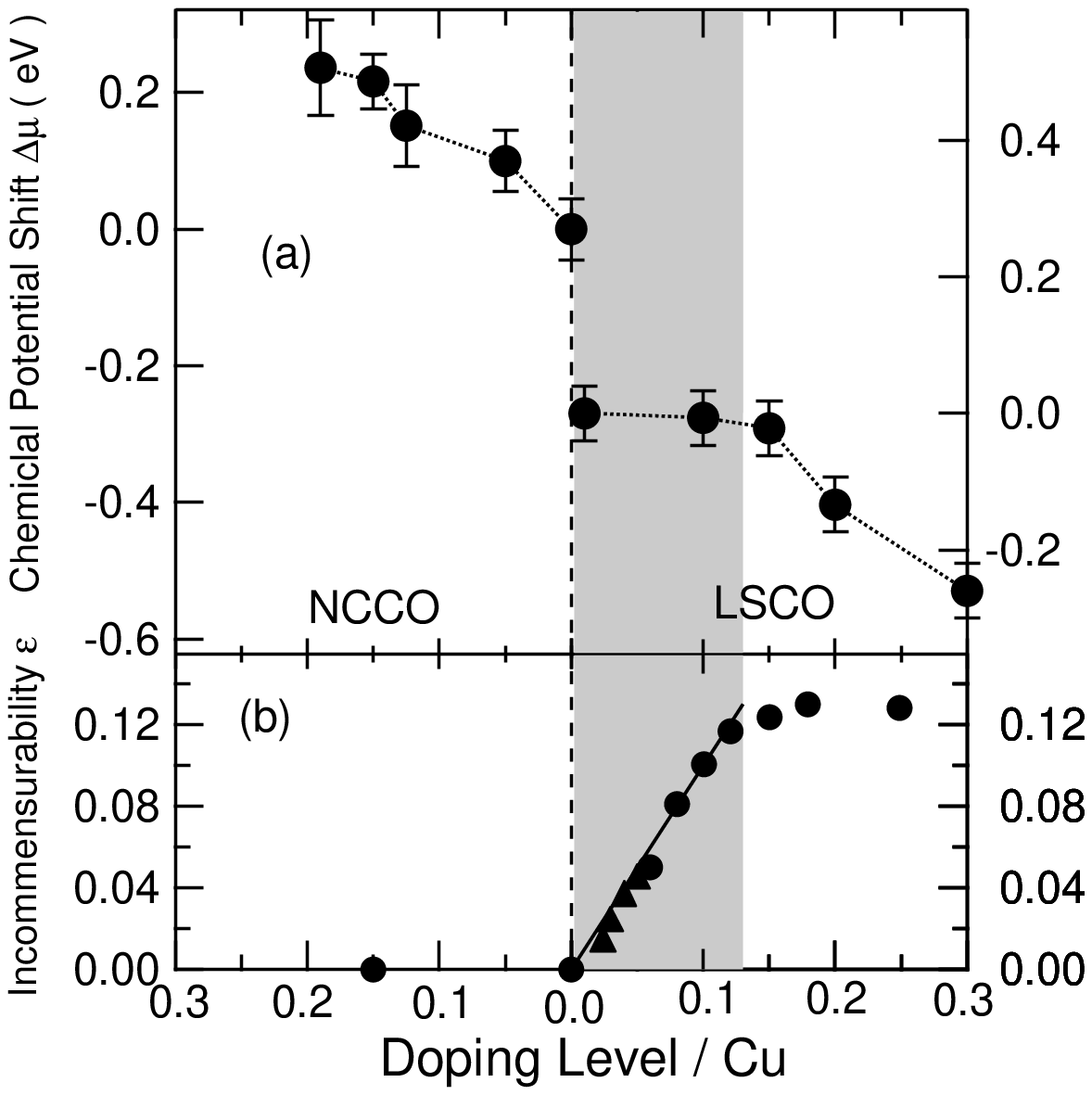}
\epsfxsize=80mm
\epsfbox{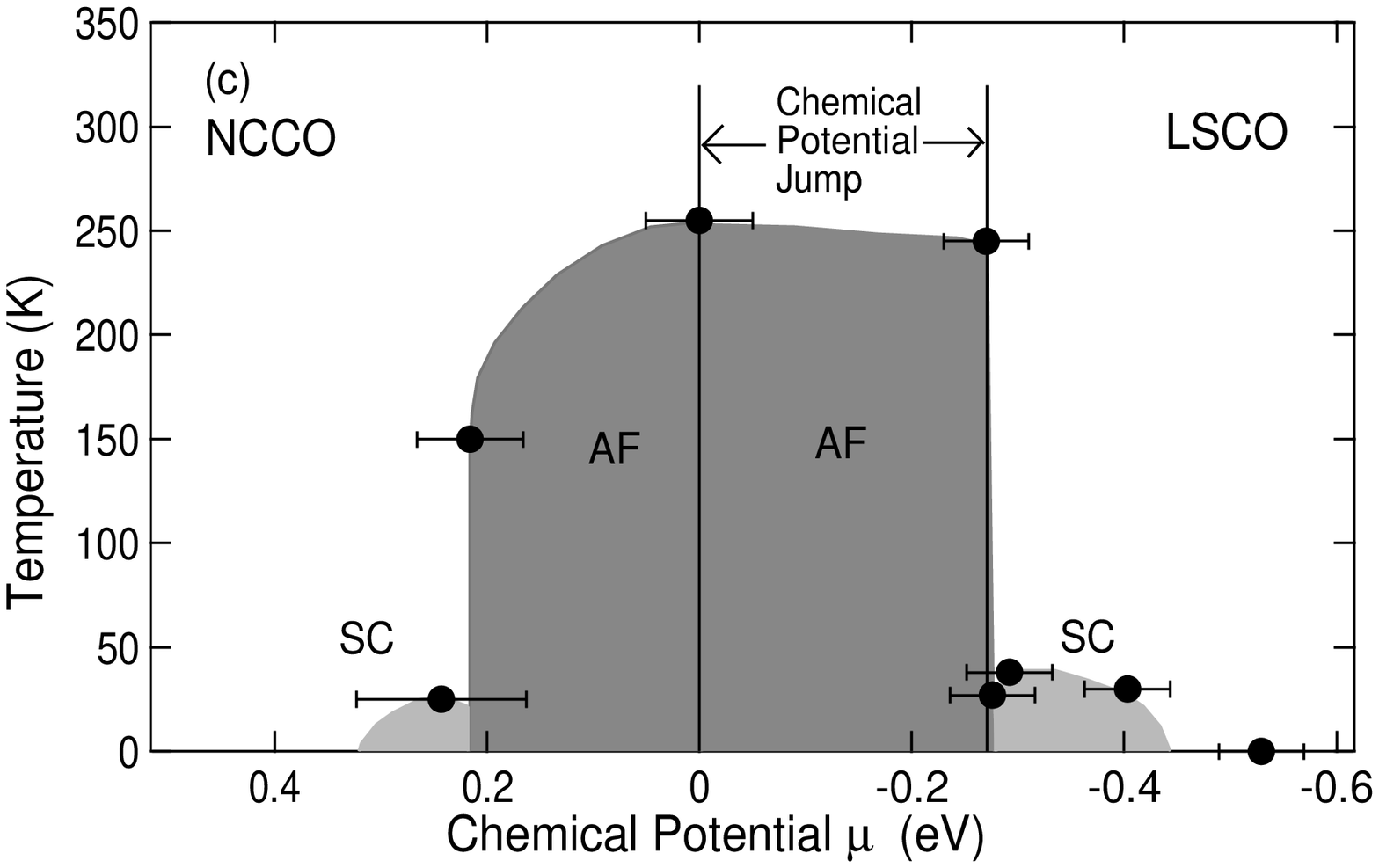}
\caption{(a): Chemical potential shift ${\Delta}{\mu}$ in NCCO and LSCO.
(b): Incommensurability ${\epsilon}$ measured by
inelastic neutron scattering experiments by 
Yamada {\it et al.} \protect\cite{LSCO_neu1,NCCO_neu}. 
In the hatched region, ${\epsilon}$ 
varies linearly and ${\Delta}{\mu}$ 
is constant as functions of doping level.
(c): ${\mu}-T$ phase diagram of NCCO and LSCO. The 
zero of the horizontal axis corresponds to the ${\mu}$ of Nd$_{2}$CuO$_{4}$.}
\end{center}
\end{figure}

Figure 3(a) demonstrates the different behaviors of ${\Delta}{\mu}$ 
between LSCO and NCCO.
In LSCO, ${\Delta}{\mu}$ is suppressed in the underdoped region
$x {\le}0.13$, whereas in NCCO ${\Delta}{\mu}$ monotonously 
increases with electron doping in the whole concentration range.
Figure 3(c) represents the phase diagram of LSCO and NCCO drawn
against the chemical potential ${\mu}$ and the temperature $T$.
One can see that the ${\mu}-T$ phase diagram is rather symmetric between the
hole doping and electron doping unlike the widely used 
${\it x}-T$ phase diagram.
That is, in both the electron- and hole-doped cases, 
the SC region is adjecent to the AF region, 
as proposed by Zhang~\cite{HTSC_phase} based on SO(5) symmetry.
The present phase diagram implies that as a function of $\mu$, $T_c$ increases 
up to the point where the superconductivity is taken over by the
AF ordering. Such a behavior is reminiscent of the superfluid-solid 
transition in the $p-T$ phase diagram of $^3$He \cite{tesa}.
\begin{figure}[ht]
\begin{center}
\epsfxsize=80mm
\epsfbox{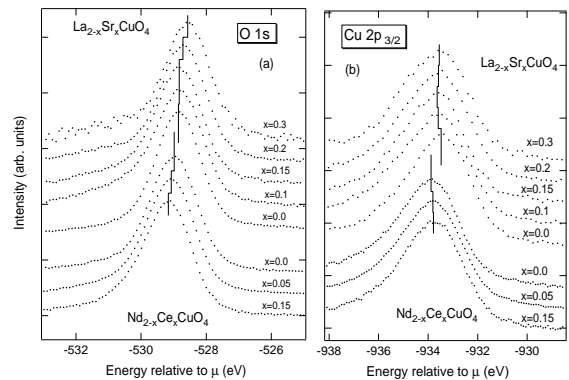}
\caption{Core-level spectra of NCCO and LSCO. (a): O 1{\it s}, (b): Cu
 2{\it p}.}
\end{center}
\end{figure}

The monotonous increase of $\mu$ in NCCO may be understood within
the simple 
rigid-band model. The chemical potential of Nd$_{2}$CuO$_{4}$
lies near the bottom of the conduction band. When electrons are doped,
${\mu}$ moves upward into this band as long as the AF ordering and hence 
the AF band structure persist as in NCCO.
This behavior is contrasted with the suppression of ${\Delta}{\mu}$
in underdoped LSCO, where the AF ordering is quickly destroyed 
and the electronic structure is dramatically reorganized by a small amount
of hole doping.
It has been suggested~\cite{LSCO_mu}
that the suppression of ${\Delta}{\mu}$ 
is related to the charge fluctuations 
in the form of stripes in LSCO. 
Indeed in La$_{2-x}$Sr$_{x}$NiO$_{4}$ (LSNO), where static 
stripe order is stable 
at $x\simeq$ 0.33~\cite{LSNO_exp}, ${\Delta}{\mu}$ is 
anomalously suppressed in the underdoped region $x{\le}0.33$~\cite{LSNO_mu}.
In LSCO, dynamical stripe fluctuations have been 
implied by inelastic neutron scattering studies~\cite{LSCO_neu1,LSCO_neu2}
whereas any sign of stripes is absent in the neutron study 
of NCCO~\cite{NCCO_neu}.
In Fig. 3, we compare the chemical potential shift 
${\Delta}{\mu}$ with the incommensurability ${\epsilon}$ which was
deduced from the neutron experiments 
both for LSCO \cite{LSCO_neu1} and NCCO \cite{LSCO_neu2}.
This figure shows that in the region 
where $\mu$ does not move
($x{\le}0.13$ in LSCO), ${\epsilon}$ linearly increases with $x$.
In NCCO, where the chemical potential ${\mu}$ monotonously moves upward,
the incommensurability does not change with $x$ (remains zero), 
in other words, static nor dynamical stripes do not exist. 
In the overdoped region of LSCO, the number of stripes saturates,  
doped holes overflow into the interstripe region and 
$\mu$ moves fast with hole doping. 

The smallness of the chemical potential jump between Nd$_{2}$CuO$_{4}$ 
and La$_{2}$CuO$_{4}$
indicates that $\mu$ lies within the CT gap
of Nd$_{2}$CuO$_{4}$ ($\sim$1.5 eV) and La$_{2}$CuO$_{4}$ ($\sim$2.0 eV).
This behavior was clearly observed in an angle-resolved photoemission (ARPES)
study of LSCO~\cite{LSCO_ARPES}, where $\mu$
is located $\sim$0.4 eV 
above the top of the valence band of La$_{2}$CuO$_{4}$.
Such a behavior is quite peculiar from the view point of 
the rigid-band model, and therefore 
a dramatic change in the electronic structure should occur with hole doping.
In NCCO, however, the 
doping-induced change is not so dramatic as in LSCO as mentioned above, 
and the chemical potential pinning well below the bottom of the 
conduction band is not very likely. 
Another possible cause of the small chemical potential jump 
is that the CT gap is indirect and is smaller 
than that estimated from the optical studies. 
In optical conductivity spectra, only the direct transition can be 
unambiguously measured, and if the gap is indirect, the gap would be estimated
much larger than the CT gap.
This idea is consistent with the recent resonant inelastic x-ray scattering 
study~\cite{RIXS} and its theoretical analysis using
$t$-$t^{'}$-$t^{''}$-$U$ model~\cite{RIXS_the}, which has yielded an 
indirect gap that is smaller than the direct one by $\sim 0.5$ eV.

In summary, we have experimentally determined the doping dependence of the
chemical potential shift ${\Delta}{\mu}$ in NCCO
and observed a monotonous shift with doping. Comparison with 
LSCO indicates that the change in the electronic
structure with carrier doping is more moderate in NCCO.  
The monotonous shift is consistent with the observation 
that spin fluctuations are commensurate in
NCCO. The small chemical potential jump between 
the $n$-type and $p$-type materials is confirmed and is 
attributed to the indirect CT gap and the chemical potential pinning 
within the CT gap in LSCO.

The authors would like to thank  S. Tesanovic, A. Ino 
and T. Mizokawa for enlightening discussions. 
Collaboration with G. A. Sawatzky and J. van Elp in the early stage 
of this work is gratefully acknowledged.
This work was supported by 
a Grant-in-Aid for Scientific Research in Priority Area
``Novel Quantum Phenomena in Transition Metal Oxides'' and a Special 
Coordination Fund for the Promotion of Science and Technology
from the Ministry of Education and Science and by
New Energy and Industrial Technology Development Organization (NEDO).


\begin{references}
\bibitem{rem}We note that the background subtraction does not affect 
the estimates of the shifts.
\end{references}
\end{document}